\documentclass[prb,twocolumn,superscriptaddress,amssymb]{revtex4}
\def\beq{\begin{equation}}
\def\eeq{\end{equation}}
\def\beqn{\begin{eqnarray}}
\def\eeqn{\end{eqnarray}}
\usepackage{graphicx}
\begin{document}
\title{Stability of the quantum spin Hall effect:\\ effects of interactions, disorder,
and $\mathbb{Z}_2$ topology}
\author{Cenke~Xu}
\affiliation{Department of Physics, University of California,
Berkeley, CA 94720}
\author{J.~E.~Moore}
\affiliation{Department of Physics, University of California,
Berkeley, CA 94720} \affiliation{Materials Sciences Division,
Lawrence Berkeley National Laboratory, Berkeley, CA 94720}
\date{\today}
\begin{abstract}
The stability to interactions and disorder of the quantum spin Hall effect (QSHE) proposed for time-reversal-invariant 2D systems is discussed.  The QSHE requires an energy gap in the bulk and gapless edge modes that conduct spin-up and spin-down excitations in opposite directions.  When the number of Kramers pairs of edge modes is odd, certain one-particle scattering processes are forbidden due to a topological $\mathbb{Z}_2$ index.  We show that in a many-body description, there are other scattering processes that can localize the edge modes and destroy the QSHE: the region of stability for both classes of models (even or odd number of Kramers pairs) is obtained explicitly in the chiral boson theory.  For a single Kramers pair the QSHE is stable to weak interactions and disorder, while for two Kramers pairs it is not; however, the two-pair case can be stabilized by {\it either} finite attractive or repulsive interactions.  For the simplest case of a single pair of edge modes, it is shown that changing the screening length in an edge with screened Coulomb interactions can be used to drive a phase transition between the QSHE state and the ordinary insulator.
\end{abstract}
\maketitle
\section{Introduction}
The possibility of an intrinsic spin Hall effect in systems with time-reversal symmetry ($T$) was first proposed for gapless semiconducting systems~\cite{murakami,sinova} and led to several experimental searches~\cite{awschalom,wunderlich}.  While a consensus has emerged that for Rashba-type spin-orbit coupling, the intrinsic effect is sensitive to disorder~\cite{vertexcorr,halperinmishchenko} (but not for p-type coupling~\cite{murakamistab,bernevigstab} in 2D or 3D), recent interest~\cite{akanegraphene}$^,$~\cite{bernevig}$^,$~\cite{qi}$^,$~\cite{onoda} has concentrated on a ``quantum spin Hall effect'' in two-dimensional systems with an energy gap in the bulk.  These models can be thought of as multiple copies of the charge Hall effect with different values of the spin, arranged so that $T$ is unbroken but the spin current is nonzero in the presence of an applied electric field: for a fixed value of the spin, $T$ is broken and there is a quantum Hall effect with zero net orbital flux~\cite{haldaneflux}.  The energy gap in the bulk means that the dominant effects of weak potential scattering are on the gapless modes at the edge of the 2D system.

An important question is whether such constructions are stable to potential scattering and other effects in real experiments.  Recently a $\mathbb{Z}_2$ topological classification~\cite{kanemele} has been proposed that is in many ways analogous to the Chern-number classification of integer quantum Hall states.  In particular, a single-particle description~\cite{kanemele} suggests that models distinguished from the insulator in the $\mathbb{Z}_2$ classification~\cite{akanegraphene,bernevig} are qualitatively more stable than other models of a two-dimensional quantum spin Hall effect~\cite{qi,onoda}.  This stability is argued on the basis of single-particle potential scattering, and an alternate form of this argument is given in Section II.  The simplest description of this $\mathbb{Z}_2$ classification is as whether the number of time-reversed pairs of gapless edge modes is even (including the ordinary insulator) or odd.  A more precise definition~\cite{kanemele} is made using the $\mathbb{Z} \times \mathbb{Z}_2$ classification of wavefunctions with a period-4 involution (time-reversal) using real K-theory.~\cite{atiyah}  This definition of the $\mathbb{Z}_2$ classification is ``topological'' (invariant under small deformations of the single-particle wavefunctions) but depends on a precise symmetry (time-reversal).

This paper clarifies by explicit calculation the nature of the difference between the two topological classes of models: the effect of the $\mathbb{Z}_2$ topological number is to restrict certain one-particle scattering processes that tend to eliminate the QSHE, but there are additional unrestricted multiparticle processes that drive the same instability and are unaffected by this conservation law.  Our emphasis is on universal properties of the effective theory; at least four~\cite{akanegraphene,bernevig,qi,onoda} specific single-particle Hamiltonians giving rise to a QSHE band structure have already been introduced, but these can be grouped into universality classes based on the number of Kramers pairs at the edge (one or two).  As expected, the $\mathbb{Z}_2$ index does lead to a larger region of stability for a single Kramers pair of edge modes than for two pairs.  However, even though for one pair repulsive interactions only reduce the stability to disorder, for two pairs repulsive interactions of moderate strength can unexpectedly stabilize the QSHE to disorder.  We compute the region of stability for the most experimentally relevant cases with and without the $\mathbb{Z}_2$ symmetry.

One way to understand the extraordinary precision of the {\it charge} quantum Hall effect is by considering possible scattering processes at the edge.  The gapless modes at the edge are determined by an integer matrix $K$ that is inherited from the Chern-Simons effective theory of the bulk condensate.  For a chiral edge (all eigenvalues of $K$ have the same sign), all low-energy excitations at the edge have the same direction of propagation: all integer quantum Hall states fall into this class.  Potential scattering cannot modify the conduction in any order of perturbation theory because there is no low-energy mode in the opposite direction into which quasiparticles can scatter.

In nonchiral edge states with modes propagating in both directions, there are allowed quasiparticle scattering processes that transfer quasiparticles from one mode to another and can reduce the edge conductance.  It was proposed by Kane, Fisher, and Polchinski~\cite{kfp} that an instability to such scattering occurs in the $\nu=2/3$ edge and is actually required to explain the observed value of the charge conductance.  The charge quantum Hall effect depend on the breaking of time-reversal symmetry, so that even in nonchiral edge states the left-moving sector of edge excitations is topologically distinct from the right-moving sector.

In contrast, the quantum spin Hall effect exists in systems with unbroken time-reversal symmetry: in this case, the left-moving sector and right-moving sector at the edge are time-reversed copies of one another.  For example, one sector may move left with spin up, while the other sector moves right with spin down.  It is then intuitively plausible that, depending on the interactions and scattering at the edge, the left-moving and right-moving sectors could scatter into each other strongly~\footnote{Note that even though pure potential scattering conserves spin, the spin-orbit coupling needed for the QSHE means that an appropriate model for the edge allows spin-flip scattering that does not break $T$.} and both localize, so that no gapless propagating modes remain and the spin conductance along the edge is zero.  This particular class of 1D localization problems with interactions and disorder is directly relevant to whether the quantum spin Hall effect is observable in experiment.

In terms of edge excitations, the $\mathbb{Z}_2$ classification found in Ref.~\onlinecite{kanemele} creates a restriction on allowed scattering processes in some ``topologically ordered'' band structures.  Our main focus in the following is on the phase diagram of the QSHE with interactions and potential scattering at the edge.  The edge theory we study in the simplest case contains two modes moving in opposite directions that are $T$ conjugates (i.e., form a Kramers pair): the two modes may carry opposite spins along some axis, so that there is a nonzero spin current in an applied electric field.  Such an edge has half as many degrees of freedom as an ordinary nonchiral Luttinger liquid with spin.  Our description of the edge is as a pair of chiral Luttinger liquids~\cite{wen1,wen2} with interactions and scattering: the kinetic term of this edge theory arises in the QSHE from a bulk theory of two Chern-Simons fields with opposite coefficients~\cite{bernevig}.  Such paired Chern-Simons theories that are $T$- and parity-symmetric have also been introduced recently for topological quantum computation.~\cite{nayakshtengel,fendleyfradkin}  The edge states of these models are similar to those discussed here and are needed in some experimental proposals~\cite{dassarma} to measure nonabelian statistics directly.  We discuss modifications required for more general cases and experimental consequences in closing.  One prediction for experiments is that by increasing the screening length of the Coulomb interaction in the case of a single Kramers pair, it is possible to drive an abrupt transition from the QSHE state to an insulator.

\section{Spinful edge excitations and time-reversal symmetry}

We start by giving an alternate explanation of the $\mathbb{Z}_2$ classification found by Kane and Mele~\cite{kanemele}, based on unbroken $T$, that is more natural for excitations of a many-particle system.  Consider the scattering induced by a time-reversal-symmetric perturbation operator $H^\prime$: effects of such scattering are expected to be significant on gapless excitations, which are localized near the edge.  Let $T$ be the time-reversal operator: then $[H^\prime,T]=0$.  $T$ is an antiunitary operator and can be decomposed into a product of a unitary operator $U$ and complex conjugation $K$: $T = U K$, with $K \psi = \psi^*$.  From this form, it is obvious that $T$ satisfies the following requirements:
\beqn \langle T\alpha|T\beta \rangle = \langle K\alpha|K\beta \rangle =
\langle\beta|\alpha\rangle \label{T1}.\eeqn For a system with angular momentum $|j,m\rangle$, $T|j,m\rangle =
i^{2m}|j,- m\rangle.$  This indicates that for bosons or for an even number of fermions, $T^2 = 1$, while for an odd number of
fermions, $T^2 = -1$ \cite{sakurai1994}.

Suppose a time-reversal-symmetric perturbation $H^\prime$ is
turned on in a 1D system with fermionic spin-half excitations.
Imagine that at $t = 0$, $n$ right-movers are excited, and consider whether they can be scattered back to $n$ left-movers by a random potential; specifically, assume that the   final state of $n$ left-movers $|\psi\rangle$ is the $T$ conjugates of the right-moving initial state $|\phi\rangle$. The matrix element for $H^\prime$ to connect these states
is \beqn \langle \psi|H^\prime|\phi\rangle &=&
\langle T\phi|H^\prime|\phi\rangle \cr &=& \langle
TH^\prime\phi|T^2\phi\rangle = (-1)^n \langle
TH^\prime\phi|\phi\rangle \cr &=& (-1)^n \langle H^\prime
T\phi|\phi\rangle = (-1)^n \langle\psi|H^\prime|\phi\rangle
\label{T2}\eeqn The second line made use of equation (\ref{T1}). For
$n$ odd, this process is forbidden because the matrix element has
to be zero. However, for $n$ even, this process can take place as
shown explicitly below.

If there is only one Kramers pair of edge modes, one moving right and one moving left with opposite spin, then the only states degenerate in energy are connected by $T$.  The above symmetry argument proves in this case that one excitation cannot be scattered by
a time-reversal-symmetric perturbation.  If there are two Kramers doublets, two degenerate states are not necessarily connected by time reversal and can be
mixed.  However, {\it two-particle} backscattering is not forbidden by $T$, even if there is only one Kramers doublet. Two right-movers excited at $t = 0$ can be mixed with two left-movers; if this backscattering is relevant, it will eliminate the QSHE as explained in the following section.

\begin{figure}
\includegraphics[width=3.5in]{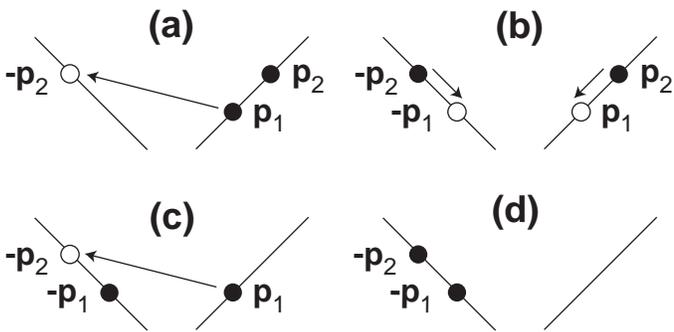}
\caption{An allowed two-particle backscattering process: (a)
particle at momentum $p_1$ scatters to intermediate state $-p_2$;
(b) particles at $\pm p_2$ interact and become intermediate
states $\pm p_1$; (c) and (d), particle at intermediate state
$p_1$ backscatters to state $-p_2$. } \label{figgs}
\end{figure}

An explicit two-particle process with a nonzero amplitude is shown in Fig.~\ref{figgs}: consider two initially right-moving particles with momenta $p_1$ and $p_2$, and allow a momentum-conserving interaction $V_{k_1,k_2 \rightarrow k_3,k_4}$ as well as potential scattering $H^\prime$.  First a particle at $p_1$ can be flipped by potential scattering to an intermediate state $-p_2$, then interact conserving momentum with the particle at $p_2$ to form another intermediate state $(-p_1, p_1)$, from which the particle of momentum $p_1$ state can be backscattered to $-p_2$.  This process has a nonzero rate in third-order perturbation theory and therefore should be included in a proper description of the edge.  The goal of the following section is to understand quantitatively how the restriction on one-particle scattering increases the stability of the QSHE.


\section{Edges with a single Kramers pair of modes}

Edges of quantum Hall systems are described by a chiral Luttinger liquid
($\chi$LL) theory related to the topological orders of the
bulk quantum Hall state.~\cite{wen1,wen2}  All edges of a certain state have the same universal
statistics matrix $K$ inherited from the bulk Chern-Simons effective theory.
The $\chi$LL action in imaginary time for a clean edge of a QH state with statistics matrix $K$ contains
$N = {\rm dim}\ K$ bosonic fields $\phi_i$ and has the form~\cite{wenrev1}
\begin{equation}
\label{action}
S_0 = {1 \over 4 \pi} \int{dx \, d\tau\, [K_{ij} \partial_x \phi_i \partial_\tau \phi_j +
V_{ij} \partial_x \phi_i \partial_x \phi_j]}
\end{equation}
where the sum over repeated indices is assumed.  $K$ is a symmetric integer matrix and $V$ a symmetric positive matrix.  $K$ gives the topological properties of the edge: the types of quasiparticles and their relative statistics.  $V$, which contains interactions (off-diagonal terms) and velocities (diagonal terms), must be positive
definite in order that the Hamiltonian be bounded below.  One major assumption in the above is that the density-density interactions are not too long-ranged, so that the spatially local form of (\ref{action}) is appropriate: changing the screening length changes the interaction matrix and in particular can be used to drive a transition in or out of the QSHE state as shown below.  The electric charges of quasiparticles are specified by an integer vector ${\bf t}$.

For the current models of the QSHE~\cite{qi,bernevig,akanegraphene}, there are only integer excitations and $K$ is a diagonal matrix of the form
\beq
K = \left(\matrix{I_n &0\cr 0&-I_n}\right)
\eeq
where $n$ is the number of time-reversed pairs of edge modes, and $I_n$ is the $n \times n$ identity matrix.  Each mode carries unit charge.  We will focus on the case $n=1$ of the graphene model~\cite{kanemele}, and the case $n=2$ that is appropriate for some semiconducting models~\cite{qi}.  Some brief statements on $n>2$ cases are also included below.

The new feature of QSHE models compared to previous work on the quantum Hall effect is that there are additional restrictions beyond charge conservation on scattering processes.
Scattering by spatially random quenched impurities is described by the action
\begin{equation}
\label{action2}
S_1 = \int{dx \, d\tau \, [\xi(x) e^{i m_j \phi_j} + \xi^*(x)
e^{-i m_j \phi_j}]}
\end{equation}
Here $\xi$ is a complex random variable and
$[\xi(x) \xi^*(x^\prime)] = D \delta(x - 
x^\prime)$, with $D$ the (real) disorder strength.
The integer vector ${\bf m}$ describes how many
of each type of quasiparticle are annihilated or created by the operator
$O_{\bf m} = \exp(i m_j \phi_j)$.  All allowed scattering operators $m_j$ are expected to appear in a disordered sample, but most of these will be
irrelevant in the RG sense as discussed in the following.  The condition for charge-neutrality is $t_i (K^{-1})_{ij} m_j = 0$.

The calculation of disorder effects in chiral Luttinger liquids differs little from the earlier work of Giamarchi and Schulz on ordinary Luttinger liquids~\cite{giamarchi}.  The first step is to determine the scaling dimension of scattering operators $O_{\bf m}$ in order to find out whether scattering grows or shrinks under rescaling.  Although the scaling dimensions are nonuniversal (i.e., depend on the nonuniversal matrix $V$), the universal $K$ matrix determines a lower bound on the scaling dimension:
\beq
\Delta_{\bf m} \geq {{\bf m}^T K^{-1} {\bf m} \over 2}.
\label{scaldimbound}
\eeq
The specific case of a null vector (${\bf m}^T K^{-1} {\bf m} = 0$) has been labeled by Haldane as a ``topological instability''~\cite{haldanetopo}, because the conditions for such an instability to exist depend on the topological $K$ matrix: the effect of such an instability is to remove a combination of modes from the low-energy theory.  The goal of the rest of the paper is to describe when such an instability eliminates the QSHE.

We first consider the case $n=1$.  The $V$ matrix can be represented in $SO(1,1)$ Lorentz coordinates~\cite{mooreedge} as
\beqn
V &=& B \left(\matrix{v&0 \cr 0& v}\right),\cr
B &=& \left(\matrix{\cosh \tau &\sinh \tau\cr\sinh \tau & \cosh \tau}\right),
\eeqn
where $v$ is a nonuniversal velocity, with $v_{\rm left}$ = $v_{\rm right}$ by $T$ symmetry, and $\tau$ parametrizes the rest of the $V$ matrix.  (The utility of this coordinate system will be clear in the $n=2$ case below.)  The scaling dimension $\Delta$ of the vertex operator $O_1 = \exp(i m_i \phi_i)$ with $m_1 = 1, m_2 = -1$ is
\beqn
2 \Delta_1 &=&  \left(\matrix{1&-1}\right) 
B \left(\matrix{1&0 \cr 0& 1}\right) 
B \left(\matrix{1\cr -1}\right)\cr
&=& 2 (\cosh 2 \tau - \sinh 2 \tau)^2.
\eeqn
However, from the above discussion of the $\mathbb{Z}_2$ classification, in a system with a single Kramers pair of edge modes the above operator is not generated.  Instead the operator $O_2$ which transfers two quasiparticles is the most relevant allowed operator, with scaling dimension $\Delta_2 = 4 \Delta_1 = 4 (\cosh 2 \tau - \sinh 2 \tau)$.  Both $\Delta_1$ and $\Delta_2$ range from 0 to $+\infty$ as $\tau$ goes from $-\infty$ to $\infty$: this is an explicit demonstration of the bound (\ref{scaldimbound}).  Note that in experimental systems the interaction is likely to be a screened Coulomb interaction, which corresponds to a negative $\tau$ as the intermode interaction strength $V_{12}=V_{21}$ is comparable to the intramode interaction strength $V_{11} = V_{22}$.

The scaling dimension of an operator determines whether that operator is
relevant in the RG sense when added to the clean action.  The operator is relevant with a spatially random coefficient when $\Delta({\bf m}) < 3/2$, as the leading-order RG flow equation for disorder strength $D$ is~\cite{giamarchi}
\begin{equation}
{dD \over d\ell} = (3 - 2 \Delta) D.
\end{equation}
Hence $O_2$ is relevant when $\Delta_2 < 3/2$, or when $\cosh 2 \tau - \sinh 2 \tau < 3/8$.  For weak disorder, there is a transition at $\tau = \tau^* = \frac{1}{2} \cosh^{-1} (73/48)$ where disorder becomes relevant and localizes both modes.  The region of stability $\tau<\tau^*$ becomes smaller when the disorder strength is finite, because
the parameter $\tau$ is increased by $D$ under RG transformations: in terms of the scaling dimension $\Delta_2$, the RG equation is~\cite{moorewen2}
\beq
{d\Delta_2 \over d\ell} = - {8 \pi \over (2 v) v^{2 \Delta_2 - 2}} {\Delta_2}^2 D.
\eeq
The Kosterlitz-Thouless-like RG flows near the critical point are similar to those found in~Ref.\onlinecite{kfp} for the $\nu=2/3$ quantum Hall state.

Since $\tau^*$ is positive, the QSHE in this model is stable to weak disorder if the intermode interaction $V_{12}=V_{21}$ vanishes, as suspected in Ref.~\cite{kanemele}.  However, one expects screened Coulomb interactions to be described by increasingly positive $\tau$ as the screening length is increased.  This suggests, first, that one should look for this type of quantum spin Hall effect in systems with strongly screened interactions, and second, that the QSHE may disappear as the screening length is increased, e.g., by changing the gates near a 2DEG.  This transition is quite sharp even if the experiment is done using current techniques where spin accumulation rather than spin currents are measured, although the amount of steady-state spin accumulation is determined by nonuniversal decay processes.
The width of the transition is set by how the localization length depends on the parameter that tunes the interactions; the Kosterlitz-Thouless RG flows imply that this length scale diverges exponentially near the transition point, so even slightly away from the transition point the localization length is quite small and the edge spin conductance will be zero.

\section{Edges with multiple Kramers pairs of modes}

We now consider the stability of the $n=2$ realizations of the QSHE.  There are 6 independent parameters of the $4 \times 4$ $V$ matrix (there would be 10 in the absence of $T$): two velocities $v_1$ and $v_2$, one intrasector rotation parameter $\theta$, two intra-Kramers-multiplet boosts $\tau^{1,2}_T$, and one inter-Kramers-multiplet boost $\tau_C$:
\beq
V = B_T B_C R \left(\matrix{v_1&0 &0&0\cr0&v_2&0&0\cr0& 0&v_2&0\cr
0&0&0&v_1}\right) R^T B_C B_T.
\eeq
Here the rotation and boost matrices are
\beqn
R &=& \left(\matrix{\cos \theta & -\sin \theta &0&0\cr\sin \theta&\cos \theta&0&0\cr0& 0&\cos \theta&-\sin \theta\cr
0&0&\sin \theta&\cos \theta}\right),\cr
B_T &=& \left(\matrix{\cosh \tau^1_T & 0 & 0 & \sinh \tau^1_T \cr
0 & \cosh \tau^2_T & \sinh \tau^2_T & 0 \cr
0 & \sinh \tau^2_T & \cosh \tau^2_T & 0\cr
\sinh^1_T & 0 & 0 & \cosh \tau^1_T}\right),\cr
B_C &=& \left(\matrix{\cosh \tau_C &0&\sinh \tau_C&0\cr
0&\cosh \tau_C&0&\sinh \tau_C\cr
\sinh \tau_C&0&\cosh \tau_C&0\cr
0&\sinh \tau_C&0&\cosh \tau_C}\right).
\eeqn
The point of this representation is that the rotation and velocity parameters do not affect scaling dimensions of vertex operators~\cite{mooreedge}.

\begin{figure}
\includegraphics[width=3.5in]{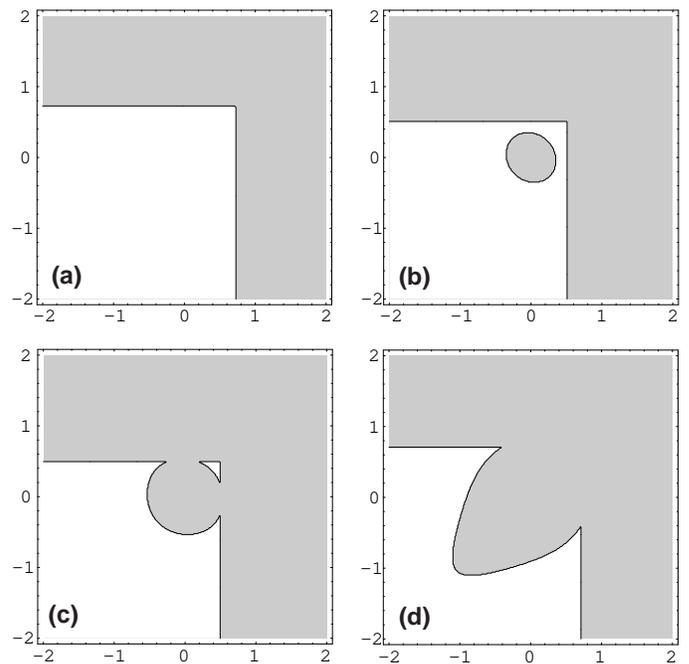}
\caption{Regions of stability (white) in the $(\tau^1_T, \tau^2_T)$ plane for values
(a) $\tau_C = -0.5$; (b) $\tau_C = -0.15$; (c) $\tau_C = -0.075$; (d) $\tau_C = 0.5$.  Note that the region of stability (b) or of instability (c) may be disconnected, and that this topological property is coordinate-independent.}
\label{phasefig}
\end{figure}

Now the lowest allowed operators either transfer one particle from a mode to the mode in the other direction that is not $T$ conjugate to the original mode, e.g. ${\bf m} = (1,0,-1,0)$, or else transfer two quasiparticles within a time-reversed pair as above.  The resulting stability criteria for scattering to be irrelevant are, for the first kind,
\beqn
&\Big[\cosh (2 \tau_C + \tau^1_T-\tau^2_T)+
\cosh (2 \tau_C + \tau^2_T-\tau^1_T) \cr &- 2 \sinh 2 \tau_C\Big]
\times \cosh (\tau^1_T+\tau^2_T) > 3,
\label{critone}
\eeqn
and for the second kind, for both branches $i=1,2$
\beq
\cosh 2 \tau_C (\cosh 2 \tau^i_T - \sinh 2 \tau^i_T) > 3/8.
\eeq
The region of stability in the $(\tau^1_T, \tau^2_T)$ plane is shown in
Fig.~\ref{phasefig} for several values of $\tau_C$.  (Obviously changes of coordinate system will change the details of these plots but not the topology.)  The region of stability is disconnected in some cross-sections at constant $\tau_C$ but is connected as a 3D object.  The prediction of this phase diagram, which is one of the main results of this paper, is that the $n=2$ QSHE can be stabilized not only for attractive interactions, as expected from the single-mode case, but even for certain choices of {\it repulsive} interactions, which are more likely to appear in experiment.

A heuristic picture of how repulsive interactions can stabilize the QSHE in this case is as follows.  With no interactions, the operators that scatter within a Kramers pair are irrelevant, but backscattering between Kramers pairs is relevant.  Turning on repulsive interactions within a pair will eventually make the intrapair scattering relevant, as in the case $n=1$ above.  However, there is an intermediate regime where repulsive intrapair interactions increase the scaling dimension of backscattering between pairs and can make it irrelevant, before backscattering within a pair becomes relevant.

This stabilization by completely repulsive interactions depends on having interpair repulsion weaker than intrapair repulsion, which is likely to be the case in real systems.  As a simple example of this stabilization, the $V$ matrix for parameters $\tau_C=0$, $\theta=0$, $v_1=v_2=1$, and $\tau^1_T = \tau^2_T = 0.485$ has all elements nonnegative:
\beq
V_s = \left(\matrix{1.50851& 0 & 0 &1.12943 \cr
0 &1.50851& 1.12943& 0\cr
0 &1.12943&1.50851&0\cr
1.12943 & 0 &0 &1.50851}\right)
\eeq

The first criterion is restrictive in that decoupled modes ($\tau^i_T = \tau_C = 0$) do not satisfy it, so that some small amount of interaction is necessary to stabilize the QSHE.  Both $n=1$ and $n=2$ models are qualitatively similar, however, in that stability to disorder is contingent upon the nonuniversal interaction strengths in the matrix $V$: in this respect both are different from the ordinary quantum Hall effect, where stability is guaranteed by the universal topological matrix $K$.

It seems possible that an instability of the second type may localize only 2 of the 4 low-energy modes (label these $1_L, 1_R, 2_L, 2_R$ according to the direction of propagation), by the following heuristic argument.  Since the eventual development of a gap via the instability is nonperturbative, this perturbative RG argument may be misleading.  When the first stability criterion (\ref{critone}) is violated, both the operator acting to pair modes $1_L$ and $2_R$ and the operator acting to pair modes $1_R$ and $2_L$ are relevant (these have the same scaling dimension by $T$), so presumably both pairs become gapped.  If the second criterion is violated but not the first, which can happen for some regions of parameter space, suppose that the time-reversed pair $1_{L,R}$ is unstable to two-particle hopping while pair 2 is stable.  This implies $\tau^1_T > \tau^2_T$.  Then the parameter $\tau^1_T$ is increased by RG transformations, which increases the left side of the first criterion (\ref{critone}) so that it remains satisfied and mode 2 can survive.

For edges with $n>2$ Kramers pairs, the parameter space is quite large: for $n=3$ there are 6 independent boost parameters, 3 velocities, and 3 rotations: 3 of the boost parameters are within Kramers pairs and 3 connect different pairs.  We can make some general comments based on the $n=2$ analysis.  Depending on the parameter values, it is possible for 0, 1, 2, or 3 of the Kramers pairs to be eliminated by scattering: operators scattering 2 quasiparticles within a Kramers pair tend to eliminate just that pair, while operators connecting two pairs tend to eliminate both pairs.  (Experimental realizations may have some additional symmetries beyond $T$ that connect the three pairs and reduce the number of free parameters; for the maximally symmetric case, either 0 or 3 pairs are eliminated).  The $n=3$ edge is actually less stable than the $n=1$ edge in that even if all the 2-particle scattering processes within a Kramers pair are irrelevant, the processes that scatter one particle from one pair to another can be relevant and drive an instability.

\section{Conclusions}

Most previous work on variants of the spin Hall effect has concentrated on noninteracting electrons.  The purpose of this paper was to understand for the quantum spin Hall effect the combined effect of interactions and disorder on the edge spin conductance.  In particular, we reviewed how the $\mathbb{Z}_2$ topological classification introduced in Ref.~\onlinecite{kanemele} leads to additional restrictions on scattering processes of edge excitations: these restrictions lead to a significantly increased region of stability of the QSHE.  The experimental conditions for the graphene case with free electrons are discussed in Ref.~\onlinecite{akanegraphene}: a prediction of our work is that the Coulomb interaction should be strongly screened for the QSHE to be realized in that system, and that reducing the screening length will drive an instability to the insulating state.  If it is someday possible to realize a QSHE of ultracold fermionic atoms, the interaction between (charge-neutral) edge excitations will be non-Coulombic and additional tuning of different instabilities may be possible.

An explicit generalization of the $\mathbb{Z}_2$ index of single-particle wavefunctions to the many-particle wavefunction is discussed briefly in Ref.~\onlinecite{kanemele}.  While the many-particle wavefunction is inaccessible in our bosonization treatment, it should be possible to confirm numerically for small systems in the clean case by modifying interaction potentials that the change in the $\mathbb{Z}_2$ index of the many-particle wavefunction occurs at the same point as the physical change in the spectrum of edge excitations.  This would be direct confirmation that the $\mathbb{Z}_2$ index remains a precisely defined concept in the presence of interactions, as is suggested by our finding that the edge structure is stable to weak interactions.


{\it Note added:} After the completion of this manuscript, we learned of independent work by Wu, Bernevig, and Zhang~\cite{wuunpub} on magnetic and nonmagnetic impurities for the $n=1$ case.
Their results for quenched nonmagnetic disorder are in quantitative agreement with ours.

\acknowledgments

J. E. M. wishes to acknowledge conversations with F. D. M. Haldane and D.-H. Lee.  The authors acknowledge support from NSF DMR-0238760 and DOE, and the hospitality of the Aspen Center for Physics (J.E.M.)
\bibliographystyle{../Newliou/apsrev}
\bibliography{../bigbib}

\end{document}